\documentclass[onecolumn]{aa}
\usepackage{graphicx}
\usepackage{txfonts}
\newcommand{\kms}{km s$^{-1}$}

\newcommand{\sig}{$\sigma$}
\newcommand{\mbh}{$M_{\rm BH}$}
\newcommand{\etal}{{\it et al.}~}
\newcommand{\chandra}{{\it Chandra}~}

\begin{document}
   \title{Black hole growth by accretion}

   \subtitle{}

   \author{Smita Mathur
          \and
          Dirk Grupe
          }

   \offprints{S. Mathur}

   \institute{Astronomy Department, The Ohio State University, 
              140 West 18th Avenue, Columbus, OH 43210, USA \\
              \email{smita@astronomy.ohio-state.edu}
             }

   \date{Received July 2004; accepted ???? 2004}

   \abstract{

   We show that black holes grow substantially by accretion at close to
   Eddington rates. Using a complete sample of soft X-ray selected AGNs,
   Grupe \& Mathur (2004) have shown that narrow line Seyfert 1
   galaxies, as a class, lie below the \mbh--\sig\ relation of normal
   galaxies. Some NLS1s, however, lie on or close to the \mbh--\sig\
   relation. Here we show that not all NLS1s accrete at close to
   Eddington rates: those with low ${\rm L/L_{Eddington}}$ are close to
   the \mbh--\sig\ relation, and those with high ${\rm L/L_{Eddington}}$
   are far. With various tests in this paper, we argue that black holes
   grow in mass substantially in their high-accretion phase and approach
   the \mbh--\sig\ relation over time. The mass growth in a low
   accretion phase, as in BLS1s and also in some NLS1s, appears to be
   insignificant. Any theoretical model attempting to explain the
   \mbh--\sig\ relation will have to explain the above observations.

   }

   \maketitle
%

\section{Introduction}

The observation of a tight correlation between the velocity dispersion
\sig\ of the the bulge in a galaxy and the mass of its nuclear
black hole \mbh\ was a surprising discovery over the last few years
(\cite{geb00a, ferr00, merr01}). Even more surprisingly, the above
relation for normal galaxies was also found to extend to active galaxies
(\cite{geb00b, ferr01}). Moreover, dead black holes were found in the
nuclei of all the observed nearby galaxies (e.g. Ho 1999). This was an
important result because it implies that nuclear activity was perhaps a
part of the life of every galaxy and that the quasar phenomenon is not just
a spectacular but cosmologically uninteresting event. A lot of
theoretical models attempt to provide explanation for the \mbh--\sig\
relation in the framework of models of galaxy formation, black hole
growth and the accretion history of active galactic nuclei
(\cite{haeh03, haeh98, adams01} and \cite{king03}). To understand the
origin of the \mbh--\sig\ relation, and to discriminate among the
models, it is of interest to follow the tracks of AGNs on the \mbh --
$\sigma$ plane.

Mathur \etal (2001) suggested that the narrow line Seyfert 1 galaxies
(NLS1s), a subclass of Seyfert galaxies believed to be accreting at a
high Eddington rate, do not follow the \mbh--\sig\ relation. [NLS1s are
defined as Seyfert galaxies with full width at half maximum of H$\beta$
lines less than 2000 \kms (\cite{ost85})]. This result was later
confirmed by Wandel (2002) and Bian \& Zhao (2003). Using a complete
sample of soft X-ray selected AGNs, Grupe \& Mathur (2004, Paper I
hereafter) determined black hole mass--bulge velocity dispersion
relation for 43 broad line Seyfert 1s and 32 narrow line Seyfert 1s. In
all the three papers listed above, the authors use luminosity and
FWHM(H$\beta$) as surrogates for black hole mass and FWHM([OIII]) as a
surrogate for the bulge velocity dispersion. Grupe \& Mathur (2004)
found that NLS1s lie below the \mbh--\sig\ relation of BLS1s, confirming
the Mathur \etal (2001) result. The statistical result was robust and
not due to any systematic measurement error. As noted by Grupe \& Mathur
(2004), this result has important consequences towards our understanding
of black hole formation and growth: black holes grow by accretion in
well formed bulges, possibly after a major merger. As they grow, they
get closer to the \mbh--\sig\ relation for normal galaxies. The
accretion is highest in the beginning and dwindles as the time goes
by. While a theoretical model to explain all the observations has yet to
come, the above result allows to rule out a class of models: e.g. the
above result does not support theories of \mbh--\sig\ relation in which
the black hole mass is a constant fraction of the bulge mass/ velocity
dispersion {\it at all times} in the life of a black hole or those in
which bulge growth is controlled by AGN feedback. A broad consistency is
found with the model of \cite{jordi03}.

At a first glance, the above result is at odds with the observation that
some NLS1s, at the low end of the observed range of velocity dispersion,
lie on/close to the \mbh--\sig\ relation (Mathur \etal (2001),
\cite{ferr01}, \cite{bian03}, and Grupe \& Mathur (2004)). As
mentioned above, the Grupe \& Mathur (2004) statistical result is
robust, in that NLS1s {\it as a class} do lie below the \mbh--\sig\
relation of normal galaxies. However, the observation of some NLS1s on/
close to the relation affects the interpretation of the result. If we
are to interpret the observations in terms of black hole growth by the
highly accreting NLS1s, why have some NLS1s already reached their
``final'' mass? In these {\it Research Notes} we propose a solution to
this apparent contradiction.


\section{The Hypothesis}

The first hint towards the resolution of the above conflict came from the
observation of Williams, Mathur \& Pogge (2004). In \chandra
observations of 17 NLS1s, they find a correlation between the soft X-ray
power-law slope $\alpha$ and the 1keV luminosity (see also
\cite{gru03a}). It has been known for many years that not all NLS1s
have steep soft X-ray spectra (\cite{bol96}). The results of Williams
\etal (2004) and Grupe \etal (2004) have shown that a significant
fraction of NLS1s have flat X-ray spectra and those with flatter spectra
are preferentially lower luminosity objects (and that absorption is not
the cause of the observed flatness of X-ray spectra in most of them).

The paradigm that NLS1s are highly accreting AGNs came from the analogy
with X-ray binaries having steep X-ray spectra in high state
(\cite{pound95}). Theoretical models of accretion disk plus corona also
confirmed that a high accretion rate relative to Eddington (\.{m}) leads
to steep soft X-ray spectra while low \.{m} accretion results in flatter
spectra (Kuraszkiewicz \etal 2000). The soft X-ray power-law slope was
found to correlate strongly with ${\rm L/L_{Eddington}}$ in Williams,
Mathur \& Pogge (2004) and in Grupe \etal (2004). The relatively flatter
spectra in some NLS1s suggest that these objects are accreting at a
substantially sub-Eddington rate, compared to the NLS1s with steep X-ray
spectra. In the framework of the black hole growth scenario of Mathur
\etal (2001) and Grupe \& Mathur (2004), these objects may then be the
ones close to the \mbh--\sig\ relation, as they would have already gone
through their high \.{m} state and their black holes have accumulated
most their mass. In the following section we test this hypothesis.

   \begin{figure} 
     \centering 
       \includegraphics[width=9cm]{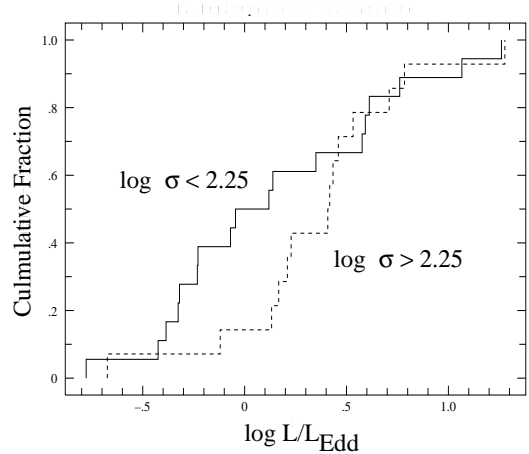}
 \caption{ Cumulative fraction for a K-S test of ${\rm L/L_{Eddington}}$
 for NLS1 with log$\sigma_{\rm [OIII]}<2.25$ (solid line) and
 log$\sigma_{\rm [OIII]}>2.25$ (dashed line). The two distributions are
 clearly different, showing that NLS1s with high ${\rm L/L_{Eddington}}$
 occupy a distinct region on the \mbh--\sig\ plane.  }
   
\end{figure}
%

\section{Tests}

If the above resolution to the black hole growth scenario is correct,
then we should find that the NLS1s close to the \mbh--\sig\ relation to
have low ${\rm L/L_{Eddington}}$ compared to those lying below the
\mbh--\sig\ relation. To test this prediction, we divided our NLS1
sample from Grupe \& Mathur (2004) in two parts, with low and high
values of \sig\, with a boundary at log$\sigma_{\rm [OIII]}$=2.25.  The
choice of the boundary came from the visual inspection of figure 1 of
Grupe \& Mathur (2004), where it was found that the NLS1s with
log$\sigma_{\rm [OIII]}$ below this value tended to be much closer to
the \mbh--\sig\ relation. Figure 1 compares the distribution of ${\rm
L/L_{Eddington}}$ for the two samples. The values of ${\rm
L/L_{Eddington}}$ are given in \cite{gru03a} and those of $\sigma_{\rm
[III]}$ are as in Grupe \& Mathur (2004, their figure 4). The
Kilogram-Smirnoff (K-S) cumulative distribution for the two samples is
significantly different with the formal K-S test probability of being
drawn from the same population P=0.1. This result is statistical in
nature. The error on values of ${\rm L/L_{Eddington}}$ for each object,
as determined in \cite{gru03a}, assuming a bolometric correction factor,
may be a factor of several. The point to note here is the {\it
difference} in the two populations with low and high \sig\ which
correspond to objects close to and away from the
\mbh--\sig\ relation respectively. Figure 1 thus shows that the objects
closer to the \mbh--\sig\ relation have lower ${\rm L/L_{Eddington}}$ and
those lying below the relation have statistically higher
${\rm L/L_{Eddington}}$.

One has to be cautious interpreting the above result, because one may
obtain high values of ${\rm L/L_{Eddington}}$ if black holes masses are
underestimated. We have emphasized in Paper I that this is not the case;
the BH masses in our sample are unlikely to be systematically
underestimated because the relationship between H$\beta$ FWHM and the
broad line region radius is well calibrated and extends to NLS1s as
well. Secondly, there is no reason for only the high \sig\ objects to
have the BH masses underestimated. Moreover, BH mass estimates using two
completely different methods give the same result: in Mathur \etal\
2001, \mbh\ was determined by fitting accretion disk models to SEDs and
in Czerny \etal\ 2001, power-spectrum analysis was used. Nonetheless
another test of the above hypothesis may be a comparison of the X-ray
power-law slopes of the two populations of high and low \sig. If our
hypothesis is correct, and if steep and flat X-ray spectra result in
NLS1s with high and low ${\rm L/L_{Eddington}}$ respectively, then we
should find that the NLS1s with low values of \sig\, i.e. those close to
the \mbh--\sig\ relation to have flatter $\alpha$ (and lower \.{m})
compared to NLS1s with high values of \sig.  In figure 2 we plot the K-S
cumulative distribution of $\alpha$ for the two populations, again using
the values from \cite{gru03a}. We find again that the two populations
are very different with the low \sig\ population having flatter spectra.
The K-S test probability of being drawn from the same population is
P=0.2.

   \begin{figure}
   \centering
   \includegraphics[width=9cm]{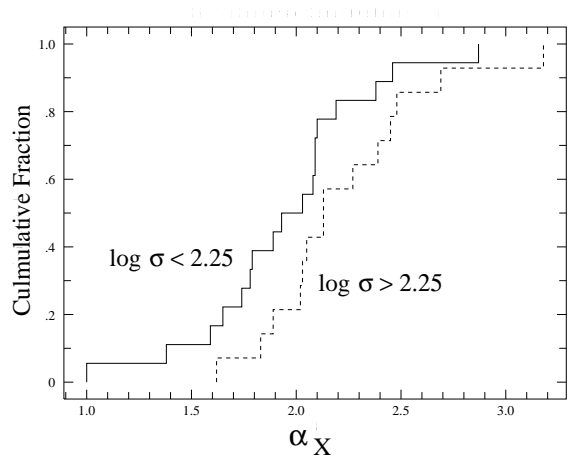}
      \caption{
Same as figure 2, but for the soft X-ray spectral index $\alpha$. Again,
the two distributions are found to be statistically different. Objects
with high ${\rm L/L_{Eddington}}$ also have steep spectra and are the
ones lying below the \mbh--\sig\ relation of normal galaxies. } 
\end{figure}
%

It is also interesting to note that the objects with high \sig\ are also
the ones with large FeII equivalent widths (figure 3). While this fact
is not directly related to the proposed BH growth hypothesis, it once
again shows that NLS1s is a mixed bag. Only NLS1s with steep X-ray
spectra appear to be those with high ${\rm L/L_{Eddington}}$ and large
FeII equivalent widths.

All these results clearly depend upon the chosen boundary between the
low and high \sig\ objects. The boundary at log$\sigma_{\rm
[OIII]}$=2.25 used above divides the total NLS1 sample of 32 into two
subsamples of 18 and 14 objects with low and high \sig\
respectively. Moving the boundary significantly either to a lower or
higher value of \sig\ would result in less than 10 objects in one data
set or the other. Nonetheless, to test the robustness of the above
results we moved the \sig\ boundary to log$\sigma_{\rm [OIII]}$=2.3
which resulted in 20 objects in low \sig\ and 12 objects in high \sig\
sets. We find that the subsamples are still different with a 
probability of being drawn from the same population P=0.2 for ${\rm
L/L_{Eddington}}$ and P=0.3 for $\alpha$. Even though the significance
of the difference goes down away from the middle boundary, the high
\sig\ objects always have preferentially high ${\rm L/L_{Eddington}}$.

As an additional test, we also determined whether the difference
$\Delta$\mbh\ between expected BH mass (as per the \mbh--\sig\ relation)
and the observed mass is correlated with ${\rm L/L_{Eddington}}$. Using
the Spearman rank correlation, we find that $\Delta$\mbh\ and ${\rm
L/L_{Eddington}}$ are correlated to better than 99.9\% significance in
the entire sample of 75 AGNs in Grupe \& Mathur (2004). If ${\rm
L/L_{Eddington}}$ is proportional to $\dot{m}$, this directly supports
the hypothesis of accretion growth of black holes.

%

   \begin{figure} 
   \centering 
   \includegraphics[width=9cm]{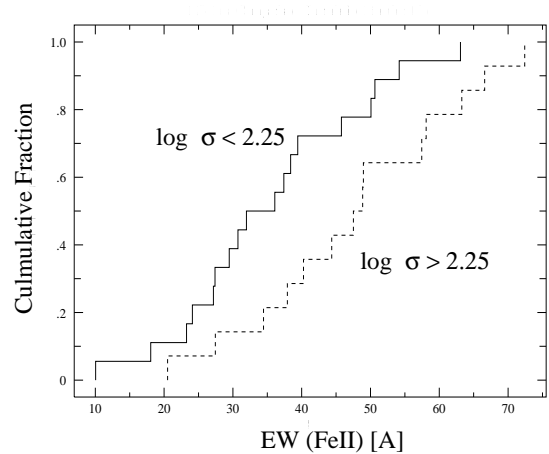}
   \caption{ Same as figure 2, but for equivalent width of FeII
   emission. Large ${\rm L/L_{Eddington}}$, steep $\alpha$ sources also
   appear to be strong FeII emitters.  } 
   \end{figure}
%

\section{Discussion}

The above tests confirm our hypothesis that NLS1s on/close to
\mbh--\sig\ relation have flatter $\alpha$ and emit at a lower fraction
of their Eddington luminosity. We emphasize again that this result is
statistical in nature, and is robust in spite of the large errors on
each of the quantities. These results have significant impact on the
NLS1 paradigm which we elaborate in Williams \etal (2004); here we
concentrate only on the implication for the black hole growth scenario.

The above hypothesis and tests support the scenario first presented in
Mathur \etal (2001) and confirmed by Grupe
\& Mathur (2004): black holes grow in mass substantially in their
high accretion phase. As they grow, they approach the \mbh--\sig\
relation for normal galaxies. The mass growth in the low accretion phase,
as in BLS1s and also in some NLS1s, appears to be insignificant. Any
theoretical model attempting to explain the \mbh--\sig\ relation will
have to explain the above observations.

Needless to say, it is vital to measure \mbh\ and \sig\ accurately to
confirm the above result. Black hole mass estimates based on H$\beta$
widths are quite secure, but the same cannot be said about estimates of
\sig\ based on [OIII] widths. Even if FWHM([OIII]) is not a good
surrogate for \sig, the nature of our result is such that
$\sigma_{[OIII]}-\sigma$ will have to be {\it different} for BLS1s and
NLS1s, and is most likely not the case as discussed in Paper
I. Moreover, there is no observational result to support such a
difference. If NLS1s had larger outflows, then they could have disturbed
their narrow lines regions more compared to BLS1s. Again, there are no
observations supporting such a case; on the contrary, absorbing outflows
are seen less often in NLS1s (Leighly 1999).  Larger ${\rm
L/L_{Eddington}}$ in NLS1s does not necessarily imply larger effective
radiation pressure. On the contrary, in objects with large soft X-ray
excesses, like NLS1s, the {\it absorbed} radiation is actually much
smaller (Morales \& Fabian 2002). There is also a general lore that highly
accreting sources with large \.{m} should have large outflows. While low
efficiency accretion must lead to outflows (as in ADIOS, Blandford \&
Begelman 1999), the same is not true for efficient accretion as in
bright Seyferts and quasars. Large outflows are observed in highly
accreting sources like broad absorption line quasars (BALQSOs), but that
depends upon the ratio of gas supply to Eddington accretion rate, and is
not inherent to the accretion process itself (R. Blandford, private
communication).

Bulge velocity dispersion is usually measured with CaII triplet line and
this technique has been used to measure \sig\ in two NLS1s (Ferrarese
\etal\ 2001). However, for many of the NLS1s in our sample, the CaII
lines fall in the water vapor band in the Earth's atmosphere. In many
NLS1s for which CaII line is accessible from ground, CaII is observed in
emission rather than in absorption (Rodriguez-Ardila \etal 2002). This
makes the use of CaII absorption features to determine \sig\ difficult
for the targets of interest. We plan to use two different methods for
alternative estimates of \sig: (1) use the CO absorption band-head at
2.29 microns to measure \sig\ directly; and (2) use high resolution
imaging of NLS1 host galaxies to measure surface brightness distribution
of bulges. One can then use fundamental plane relations to determine
\sig. Alternatively, we will determine the bulge luminosities and find
the locus of NLS1s on the \mbh-L$_{Bulge}$ relation.  Once again, the
objective is to find out if there exists a statistical difference in the
relation between black hole mass and bulge luminosity for the two
populations of BLS1s and NLS1s. We plan to use all these methods to
determine the locus of highly accreting AGNs on the \mbh--bulge
relations and so fully understand the role of accretion on black hole
growth.

\begin{acknowledgements}
      
\end{acknowledgements}


\begin{thebibliography}{}
\bibitem[Adams et al. 2001]{adams01} Adams, F.C., Graff, D.S., \& Richstone,
D.O. 2001, \apj, 551, L31
\bibitem[Bian \& Zhao 2003]{bian03} Bian, W., \& Zhao, Y., 2003, MNRAS in
press, astro-ph/0309701
\bibitem[Blandford \& Begelman 1999]{bb99} Blandford, R. \& Begelman,
 M. 1999, MNRAS, 303, L1
\bibitem[Boller et al. 1996]{bol96} Boller, T., Brandt, W.N., \& Fink, H.H.,
1996, \aap, 305, 53
\bibitem[Ferrarese \& Merritt 2000]{ferr00} Ferrarese, L., \& Merritt, D.,
2000, \apj, 539, L9
\bibitem[Ferrarese et al. 2001]{ferr01} Ferrarese, L., Pogge, R.W., Peterson,
B.M., Merritt, D., Wandel, A., \& Joseph, C.L., 2001, \apj, 555, L55
\bibitem[Gebhardt et al. 2000a]{geb00a} Gebhardt, K., Bender, R., Bower, G.,
Dressler, A., Faber, S.M., et al., 2000, \aap, 539, L13
\bibitem[Gebhardt et al. 2000b]{geb00b} Gebhardt, K., Kormendy, J., Ho, L.C.,
Bender, R., Bower, G., et al., 2000, \apj, 543, L5
\bibitem[Grupe et al. 2004]{gru03a} Grupe, D., Wills, B.J., Leighly, K.M., \&
Meusinger, H., 2004, \aj, 127, 1799
\bibitem[Grupe \& Mathur 2004]{gm04} Grupe, D. \& Mathur, S. 2004,
 ApJL, 606, 41 (Paper I)
\bibitem[Haehnelt 2003]{haeh03} Haehnelt, M., 2003, 
Classical and Quantum Gravity, 20, S31
\bibitem[Haehnelt et al. 1998]{haeh98} Haehnelt, M.G., Natarajan, P.,
 \& Rees,M.J., 1998, \mnras, 300, 817
\bibitem[Ho 1999]{ho99} Ho, L.C., 1999, in
'Observational Evidence for the Black Holes in the Universe, p157 
\bibitem[King 2003]{king03} King, A., 2003, ApJL, 596, L27 
\bibitem[Kuraszkiewicz et al. 2000]{joanna00} Kuraszkiewicz, J.,
 Wilkes, B., Czerny, B., \& Mathur, S. 2000, ApJ, 542, 692
\bibitem[Leighly 1999]{lei99} Leighly, K.M., 1999, ApJS, 125, 317 
\bibitem[Mathur 2000]{mat00} Mathur, S., 2000, \mnras, 314, L17
\bibitem[Mathur et al. 2001]{mat01} Mathur, S., Kuraszkiewicz, J., \& Czerny,
B., 2001, New Astronomy, Vol. 6, p321
\bibitem[Merritt \& Ferrarese 2001]{merr01} Merritt, D., \& Ferrarese, L.,
2001, \apj, 547, 140
\bibitem[Miralda-Escud\'e \& Kollmeier (2004)]{jordi03} 
Miralda-Escud\'{e}, J. \& Kollmeier, J.A.,  2004, \apj, submitted.
\bibitem[Morales \& Fabian 2002]{mf02} Morales, R. \& Fabian, A. 2002
 MNRAS 329, 209
\bibitem[Osterbrock \& Pogge 1985]{ost85} Osterbrock, D.E., \& Pogge, R.W.,
1985, \apj, 297, 166
\bibitem[Pounds et al. 1995]{pound95} Pounds, K.A., Done, C., \& Osborne, J.,
1995, \mnras, 277, L5
\bibitem[Rodriguez-Ardila \etal\ 2002]{rod02} Rodriguez-Ardila, A.,
  Viegas, S.M.  , Pastoriza, M.G., \& Prato, L., 2002, ApJ, 565, 140
\bibitem[Wandel 2002]{wan02} Wandel, A., 2002, \apj, 565, 762
\bibitem[Williams, Mathur \& Pogge 2004]{rik04} Williams, R., Mathur,
 S. \& Pogg e, R., 2004, ApJL, in press.
\end{thebibliography}
\end{document}